\documentclass[preprint,3p,twocolumn]{elsarticle}

\usepackage{amsmath}
\usepackage{txfonts}
\usepackage[protrusion=true,expansion=true]{microtype}
\usepackage{hyperref,xcolor}
\usepackage[T1]{fontenc} 
\usepackage{graphicx}
\usepackage{ulem}
\usepackage{tikz}
\usepackage{enumitem}
\usepackage{color} 
\usepackage{array,multirow,makecell}

\graphicspath :
\graphicspath{{Figures/}}

\begin{document}

\begin{frontmatter}

\title{Life Cycle Analysis of the GRAND Experiment}

\author[a,b]{Leidy T. Vargas-Ib\'a\~nez}
\author[c,d]{Kumiko Kotera}%
\ead{kotera@iap.fr}

\author[e]{Odile Blanchard}
\author[a]{Peggy Zwolinski}

\author[f]{\\Alexis Cheffer}
\author[f]{Mathieu Collilieux}
\author[f]{Paul Lambert}
\author[f]{Quentin Lef\`ebvre}
\author[f]{Thomas Protois}

\address[a]{%
 Universit\'e Grenoble Alpes, CNRS, Grenoble INP, G-SCOP, 38031, Grenoble CEDEX 1, France }
 \address[b]{IFP Energies Nouvelles, 92852, 1-4 Av. Bois Pr\'eau , Rueil Malmaison, France.}
\address[c]{
Sorbonne Universit\'e, CNRS, UMR 7095, Institut d'Astrophysique de Paris, 98 bis boulevard Arago, 75014 Paris, France}
\address[d]{Vrije Universiteit Brussel, Physics Department, Pleinlaan 2, 1050 Brussels, Belgium}
\address[e]{
 Université Grenoble Alpes, GAEL Laboratory (UMR CNRS 5313, UMR INRAe 1215), CS 40700 - 38058 Grenoble CEDEX 9, France
}%

\address[f]{Grenoble INP - Ense3, UGA, France}


\begin{abstract}
The goal of our study is to assess the environmental impact of the installation and use of the Giant Radio Array for Neutrino Detection (GRAND) prototype detection units, based on the life cycle assessment (LCA) methodology, and to propose recommendations that contribute to reduce the environmental impacts of the project at later stages. The functional unit, namely the quantified description of the studied system and of the performance requirements it fulfills, is to detect radio signals autonomously during 20 years, with 300 detection units deployed over 200\,km$^2$ in the Gansu province in China (corresponding to the prototype GRANDProto300). We consider four main phases: the extraction of the materials and the production of the detection units (upstream phases), the use and the end-of-life phases (downstream phases), with transportation between each step. An inventory analysis is performed for the seven components of each detection unit, based on transparent assumptions. Most of the inventory data are taken from the Idemat2021 database (Industrial Design \& Engineering Materials). Our results show that the components with the highest environmental impact are the antenna structure and the battery. The most pregnant indicators are `resource use', mineral and metals'; `resource use, fossils'; `ionizing radiation, human health'; `climate change'; and `acidification'. Therefore, the actions that we recommend in the first place aim at reducing the impact of these components. They include limiting the mass of the raw material used in the antenna, changing the alloy of the antenna, considering another type of battery with an extended useful life, and the use of recycled materials for construction. As a pioneering study applying the LCA methodology to a large-scale physics experiment, this work can serve as a basis for future assessments by other collaborations.
\end{abstract}


\end{frontmatter}


\section{Introduction} 

With the growing concern over climate and the environment, the scientific community has started to assess the environmental impact of its activity and infrastructure (e.g., \cite{Mariette2022}, see also\footnote{\href{https://labos1point5.org/les-colloques}{https://labos1point5.org/les-colloques}} and references therein, and for studies specific to the astronomy and astrophysics communities: \cite{matzner2019astronomy,stevens2019imperative,barret2020estimating,2020NatAs...4..816F,Jahnke_2020,Portegies_Zwart_2020,Burtscher_2020,Stevens2020,2021NatAs...5..857B,2021NatAs...5..861A,Aujoux2021,2021NatRP...3..386A,vandertak2021,2021JATIS...7a7001F,2022NatAs...6..503K}).
In order to make scientific research environmentally sustainable, it appears necessary to challenge and rethink the way research is being conducted.

Physics and astrophysics communities are often built around large-scale experimental projects. Evaluating the environmental impacts of these projects at an early phase (laboratory or prototype) of research and development would allow the proposition of alternatives that redirect technology towards lower environmental impact levels \cite{Tsoy2020}. The impact of the project, including the hardware equipment, and all aspects of collaboration research (travel, digital) can be calculated in terms of carbon footprint \cite{Aujoux2021,2021NatRP...3..386A,2020NatAs...4..816F,2021JATIS...7a7001F,2022NatAs...6..503K}.
Its environmental impact may also be assessed using the life cycle assessment (LCA) methodology. 
However, to our knowledge the environmental impacts of a physics experiment have not yet been reported in the literature using the LCA methodology.

Here we present a simplified LCA of the Giant Radio Array for Neutrino Detection (GRAND) experiment. GRAND is a planned large-scale observatory of ultra-high-energy cosmic particles — cosmic rays, gamma rays, and neutrinos with energies exceeding $10^8\,$GeV  \cite{GRAND20}. GRAND is in its prototyping phase, with the imminent deployment of the first antennas of GRANDProto300: 300 antennas will be installed over a desert area of $200\,{\rm km}^2$ in the Gansu province in China, and will serve as a test bench for the further steps of the experiment. The Collaboration consists today of more than 90 international researchers and engineers. Based on the output of GRANDProto300, 10\,000 antennas will be designed and deployed by 2028 over 10\,000 km$^2$, likely extending the same location in China. This will build the next phase of GRAND, GRAND10k, and be the first large-scale sub-array of the experiment. The ultimate design of GRAND will duplicate GRAND10k at $10-20$ different radio-quiet locations worldwide, to achieve a total of 200\,000 antennas over 200\,000\,km$^2$ in the  2030s (GRAND200k).

The GRAND project was the subject of the pioneering study on the carbon footprint of a large-scale experiment \cite{Aujoux2021,2021NatRP...3..386A}. It was found that, beyond the small-scale prototyping stage, hardware equipment had an equivalent or dominant carbon footprint, compared to the other two major emission sources considered (travel and digital technologies). This result called for a more detailed analysis of the environmental impacts of the hardware equipment, and triggered the present work.

The goal of this study is to assess the environmental impact of the installation and use of the GRAND prototype detection units, based on the LCA methodology, and to propose recommendations that contribute to reduce the environmental impacts of the project at later stages. 
Due to the international character of this experiment, the LCA allows us to develop the project considering the environmental commitments of the countries involved \cite{DONG2021,COP26}. 
This study is timely, as the project is still in its prototyping stage, and the GRAND Collaboration will carefully weigh the environmental impacts estimated in this work in the decision making for the design of the next experimental phases.

As a pioneering study applying the LCA methodology to a large-scale physics experiment, this work can serve as a basis for future assessments by other collaborations. Our calculation is fully transparent and based on open source data, in particular on the LCA simplified  database Idemat2021 \cite{DelfUniversityofTechnology}. \\

Section~\ref{section:methodology} presents an overview of the LCA methodology, the definition of the functional unit and the system boundaries for our study. Section~\ref{section:inventory} provides the inventory for the different components of the GRAND prototype detection units. Section~\ref{section:results} presents and discusses the results of the simplified LCA. Action plans are proposed in Section~\ref{section:action_plans} and conclusions and perspectives are presented in Section~\ref{section:conclusions}.

\section{LCA methodology} \label{section:methodology}

There exist different techniques and methods to assess the environmental aspects and potential impacts associated with a product or service throughout its entire life cycle  \cite{Morgan2012, Payraudeau2005,Kim2019}. In this study, a simplified LCA is used to conduct the analysis of the environmental impact assessment, following the ISO 14040:2006 and ISO 14044:2006 \cite{ISO,ISO2}. This method takes into account all the mass and energy flows (inputs and outputs) associated with the life cycle of a product, to quantify its potential environmental impacts. The LCA methodology is iterative, and involves the following steps: goal and scope definition, inventory analysis, impact assessment and interpretation \cite{ISO2}. 

Our final goal is to identify the best actions to improve the environmental performance from a perspective “cradle-to-grave”.

\subsection{Goal and scope}

The main objective of this study is to estimate the life cycle environmental impacts of the detection units of the 300-antenna prototype of GRAND, GRANDProto300. 
The results will help to identify hotspots and key improvement areas for the next phases of the experiment GRAND10k and GRAND200k.

\subsubsection{Functional unit}

The functional unit (FU) is a quantification of the identified functions of a product. Its main purpose is to provide a reference base on which input and output data (mass and energy balances) are normalised. It must be measurable, clearly defined and consistent with the objective and scope of the research \cite{ISO,ISO2}.

Considering the objective of this study, the functional unit is to detect radio signals autonomously during 20 years, with 300 detection units deployed over a surface of 200\,km$^2$. The impact of digital technologies (related for example to data handling) and of travel for members of the GRAND Collaboration, are not taken into account in this LCA, as their impacts will remain the same, whatever the options are for the detection units.

The next phases of the experiment will not be considered in this study. However, the GRAND10k experiment will build on the results of the LCA of GRANDProto300.

The multiple sub-arrays of the GRAND200k phase will be located in several locations worldwide (South America, South Africa, Australia, China/Mongolia/Kazakhstan are considered as serious options). The projection for the GRAND200k phase is hence less straightforward. 
It will be discussed in Section~\ref{section:conclusions}. %

\subsubsection{System boundaries}

\begin{figure*}[!tb]
    \centering
    \includegraphics[width=0.8\textwidth]{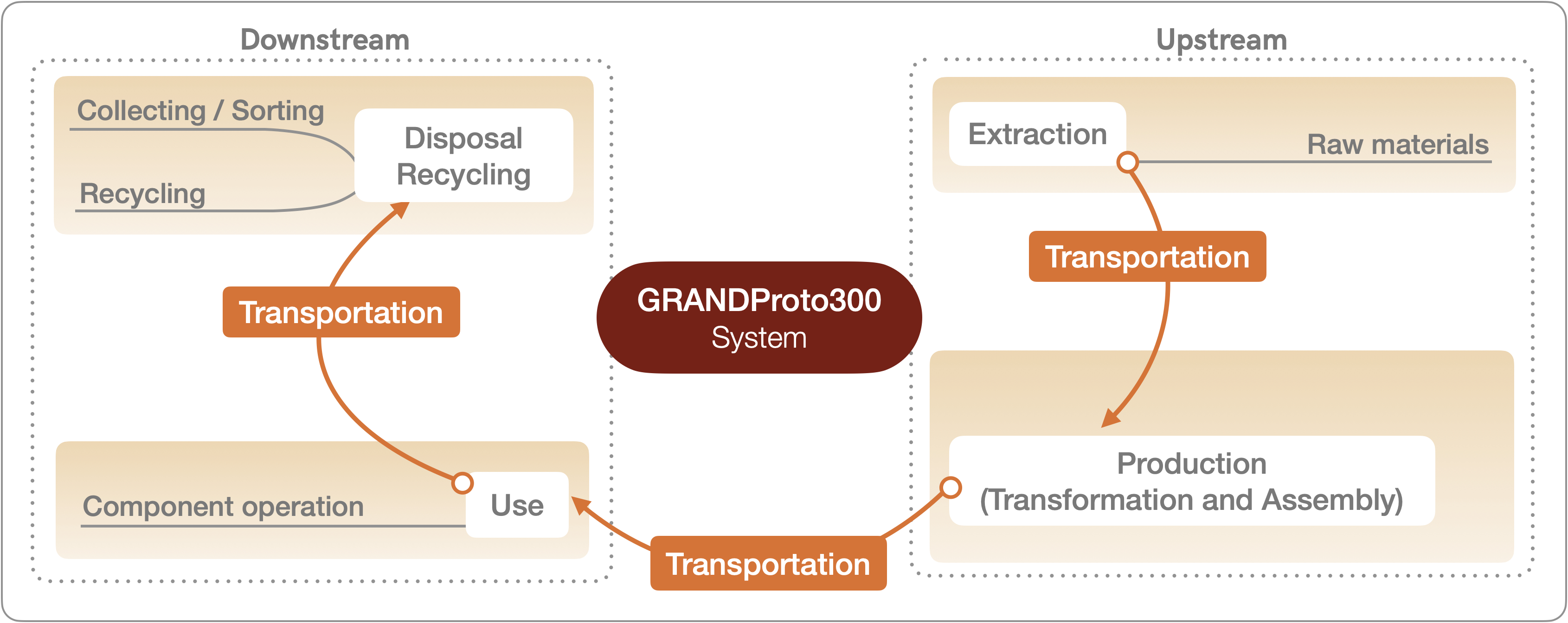}
    \caption{System boundaries. This work considers 4 main phases in the LCA of the GRANDProto300 project: the extraction of the materials and the production of the detection units (upstream phases), the use and the end-of-life phases (downstream phases), with transportation between each step.} 
    \label{fig:boundaries}
\end{figure*}

The establishment of system boundaries is crucial as it determines the unit processes to be included in the life cycle analysis. In this research we consider 4 main phases in the life cycle of the GRANDProto300 project: the extraction of the materials and the production of the detection units (upstream phases), the use and the waste treatment phases (downstream phases), with transportation between each step. The system boundaries over which the inventory is performed in this study are presented in Fig.~\ref{fig:boundaries}. We evaluate the different phases of the life cycle for seven components of the system: the antenna structure, the battery, the photovoltaic panel, the electronic card, power electronics, the GPS antenna, and the wifi antenna. 

Differents impacts were not taken into account, given the exploratory nature of this R\&D project :
\begin{itemize}
    \item The impacts of the infrastructure, transformation and the people who perform the work (travel, transportation and everything related to the work of the employees) 
    \item The impacts of the ancillary services (such as assembly, testing of equipment, maintenance or disassembly) 
    \end{itemize}

The extraction and transformation of the different materials required for the production of the seven components analyzed in this study are taken into account. However some specific hypotheses, based on the information available in the Idemat 2021 database, are presented in section ~\ref{section:inventory}.  

Consequently, the results presented in the paper are only lower limits to the true environmental impacts. But they will be useful in addressing the design issues raised by these systems.

\subsection{Life cycle inventory (LCI)} 

To conduct a LCA, it is necessary to quantify the relevant inputs and outputs in the product system. This iterative process involving data collection and calculation procedures is called inventory analysis \cite{ISO,ISO2}. In this study, most of the inventory data are taken from the Idemat2021 database (Industrial Design \& Engineering Materials) \cite{DelfUniversityofTechnology} from the Delft University of Technology. This free and open source database is a compilation of LCI data of the Sustainable Impact Metrics Foundation, SIMF. It provides the environmental impacts for different impact categories of materials, industrial processes, transportation and recycling.

In this research, data from the same database are primarily used to decrease the sources of uncertainty, since each database may contain different inventories and calculation methods \cite{Chen2021}.  Nevertheless, as a last resort, if we cannot find the information that we need in our database, we use coefficients from other studies. The inventory of the GRAND experiment is further discussed in Section~\ref{section:inventory}

\subsection{Life cycle impact assessment (LCIA)}
There are different methods for calculating the life cycle impact assessment \cite{Althaus2004}, among which two mainstream ways to derive characterisation factors: midpoint level and endpoint level \cite{report,Dong2014}. The midpoint indicators allow quantifying the overall effects of the substances emitted or consumed, by grouping the inventory with similar effects. Whereas, the endpoint indicators focus on quantifying macro changes in ecosystems, at the end of the cause-effect chain \cite{Bare2000}.

In this work, the midpoint approach is used because it allows to analyse the impacts that appear in the middle of the causal chain (CO2 emissions for example) rather than final and global damages that include several impacts (biodiversity loss for example). Therefore, the midpoint approach allows a more detailed environmental impact analysis of the different project stages, taking into account the information of the different physical and material flows. Twelve midpoint categories of environmental impacts are selected to assess the total impact of the project:

\begin{itemize}
\item Climate change 
\item Ozone depletion 
\item Photochemical ozone formation 
\item Acidification 
\item Freshwater eutrophication 
\item Marine eutrophication
\item Particulate matter 
\item Land use 
\item Resource use, fossils 
\item Resource use, minerals and metals
\item Terrestrial eutrophication
\item Ionizing radiation, human health
\end{itemize}

They are selected based on well-known LCA calculation methods (see Table~\ref{table:impacts}).

\begin{table*}[!tb]
    \centering
    \includegraphics[width=0.95\textwidth]{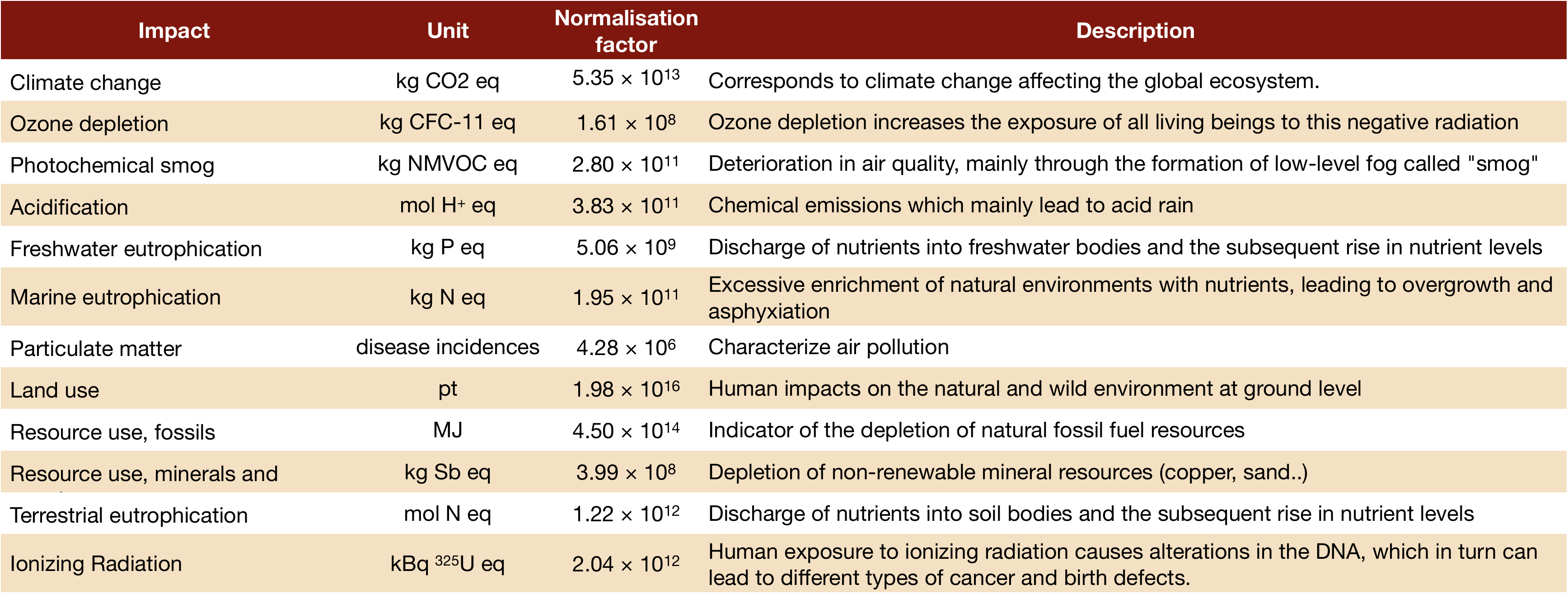}
    \caption{Description of the midpoint categories considered in this study and their corresponding normalisation factors from \cite{Sala2018}. Units are detailed in~\ref{app:units}}
    \label{table:impacts} 
\end{table*}

\section{Inventory of the GRAND experiment}\label{section:inventory}

\begin{figure}[h]
    \centering
    \includegraphics[width=0.49\textwidth]{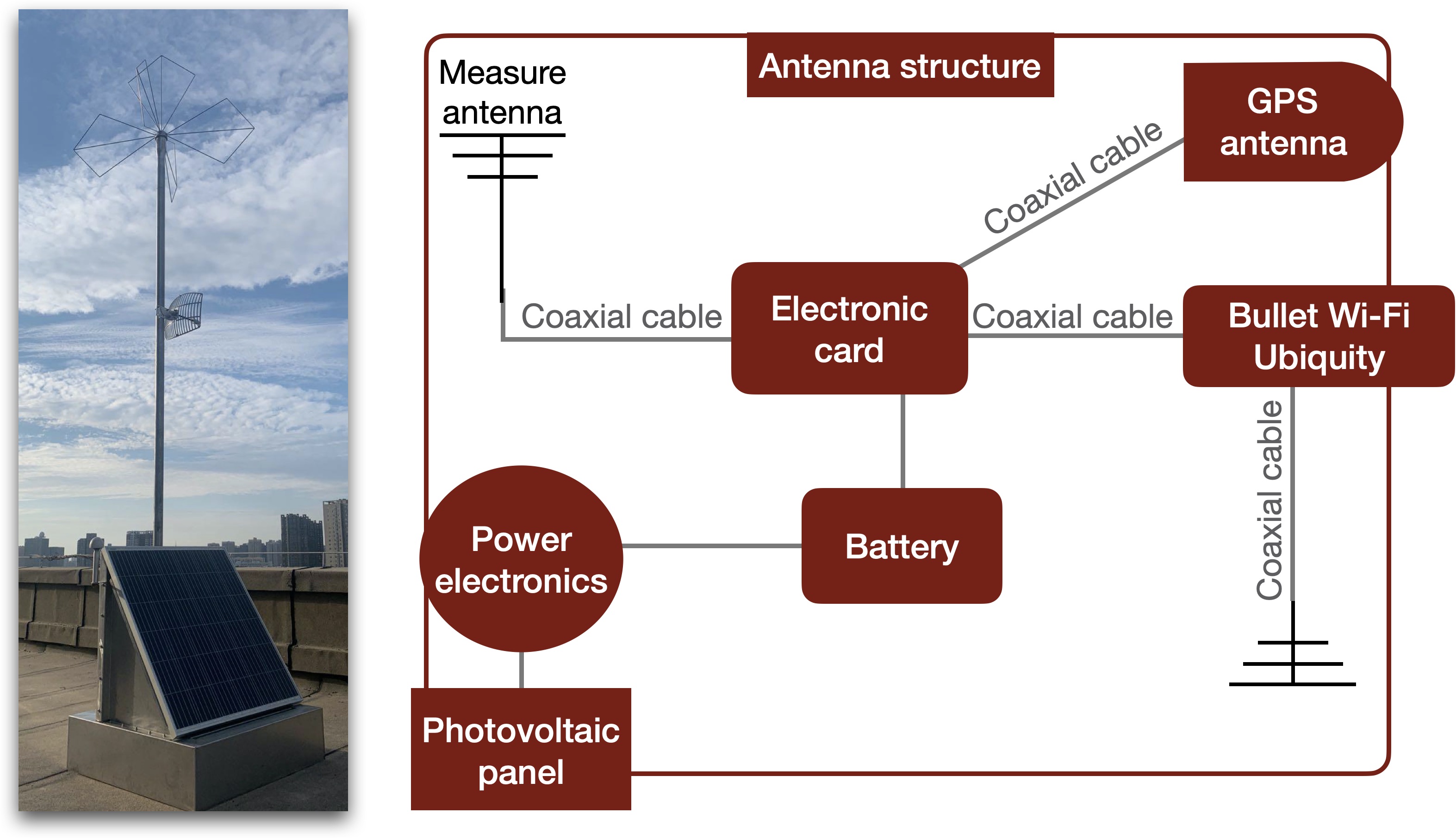}
    \caption{{\it Left:} Picture of a GRAND prototype (GRANDProto300) detection unit. {\it Right:} Sketch of the 7 components of the GRANDProto300 detection unit.}
    \label{fig:inventory}
\end{figure}

The GRAND prototype detection units, pictured in Fig.~\ref{fig:inventory}, are made up of 7 main components. An inventory analysis is performed for each of these components.
Some elements of the detection unit, such as the antenna nut (which connects the arms to the antenna pole and to the electronic card via cables) or the connectors have been neglected, because of their low impact compared to the 7 components treated in the study.

In this section, we first detail general assumptions concerning raw materials, recycling and transportation. 
We then overview the constituents and assumptions used to perform the inventory of the 7 components. 

\subsection{Assumptions on raw materials and end-of-life}\label{section:assumptions_recycling} 

Regarding the raw materials used, it is assumed that, unless otherwise specified, {\it primary} materials are always used for the different components of the GRAND project. For example, aluminium is assumed to be obtained from bauxite, not from aluminium scraps.

For the life cycle, it is assumed that all the recyclable, reusable materials are fully recycled or upcycled. The recycling percentage for each material is established based on the availability of data in the Idemat database. 

In this work, the end-of-life of materials and equipment is modeled considering the recycling and upcycling activities available in the Idemat database. For materials for which end-of-life information is not available in Idemat, it is assumed that waste management is performed by incineration/landfill (considering 50\% incineration and 50\% landfill), for which a general and average impact factor is obtained from the Ecoinvent database \cite{Ecoinvent2021}.

While we consider waste treatment of the equipment materials using the available activities in Idemat and Ecoinvent database \cite{Ecoinvent2021}, we do not consider process-specific waste treatment requirements due to a lack of data, such as additional energy demands for the dismantling of the different components.

\subsection{Assumptions on transportation}
\label{section:assumptions_transport} 

Figure~\ref{fig:boundaries} shows that 3 types of transportation are taken into account: i) transportation between the extraction and production sites, ii) 
transportation between production and operation sites and iii) transportation between the operation site and the disposal or recycling sites. We call i) upstream transportation and ii) upstream to downstream transportation, while iii) is called downstream transportation. 
The different transportation distances are shown in Table~\ref{table:inventory_main}; the detail of the transport values can be found in the Supporting Information\footnote{\label{excel} 
\href{https://grand.cnrs.fr/wp-content/uploads/2023/07/LCA_GP300-_23_06_2023.xlsx}{https://grand.cnrs.fr/wp-content/uploads/2023/07/LCA\_GP300-\_23\_06\_2023.xlsx}.}
.

The following assumptions are made to calculate the transportation inventory:

\begin{itemize}
\item The production of the antennas will take place in Xi'An, China, in a plant located 1900\,km away from the operation site. The operation site is located in the West of the Gansu province. The disposal and recycling site factories are assumed to be in Xi'An. Other recycling centers and factories exist closer to the GRANDProto300 deployment site. This improved scenario will be discussed in Section~\ref{section:action_plans}.

\item The detection units will be transported from Xi'An to Gansu, and then transported back to the disposal or recycling site, as all parts of the system.

\item 
In China, rail and road transportation are equally used\footnote{\href{https://www.statista.com/statistics/264809/freight-traffic-in-china/}{https://www.statista.com/statistics/264809/freight-traffic-in-china/}}. However, the Idemat database provides train coefficients only within the USA. Therefore, we choose to restrict all national transportation in China to truck transportation.

\item Among the options for road transportation provided in the Idemat database, we rule out the one with European norms (`Truck + trailer Euro 6'), which is not adapted to the project location. The option `Truck + trailer 24 tons net' in the Idemat database is chosen, as it provides an upper limit to the calculated values. 

\end{itemize}

\subsection{Assumptions on each of the 7 components}\label{section:assumption_components}

In this section, we detail  the assumptions made on the 7 components of the inventory, sketched in Fig.~\ref{fig:inventory}, for the upstream and downstream phases.  
Table~\ref{table:inventory_main} presents the life cycle inventory. Only the most representative materials and transformation activities are shown. The detail of the inventory can be found in the Supporting Information (see Footnote~\ref{excel}).

\begin{table*}[!tb]
    \centering
    \includegraphics[width=0.95\textwidth]{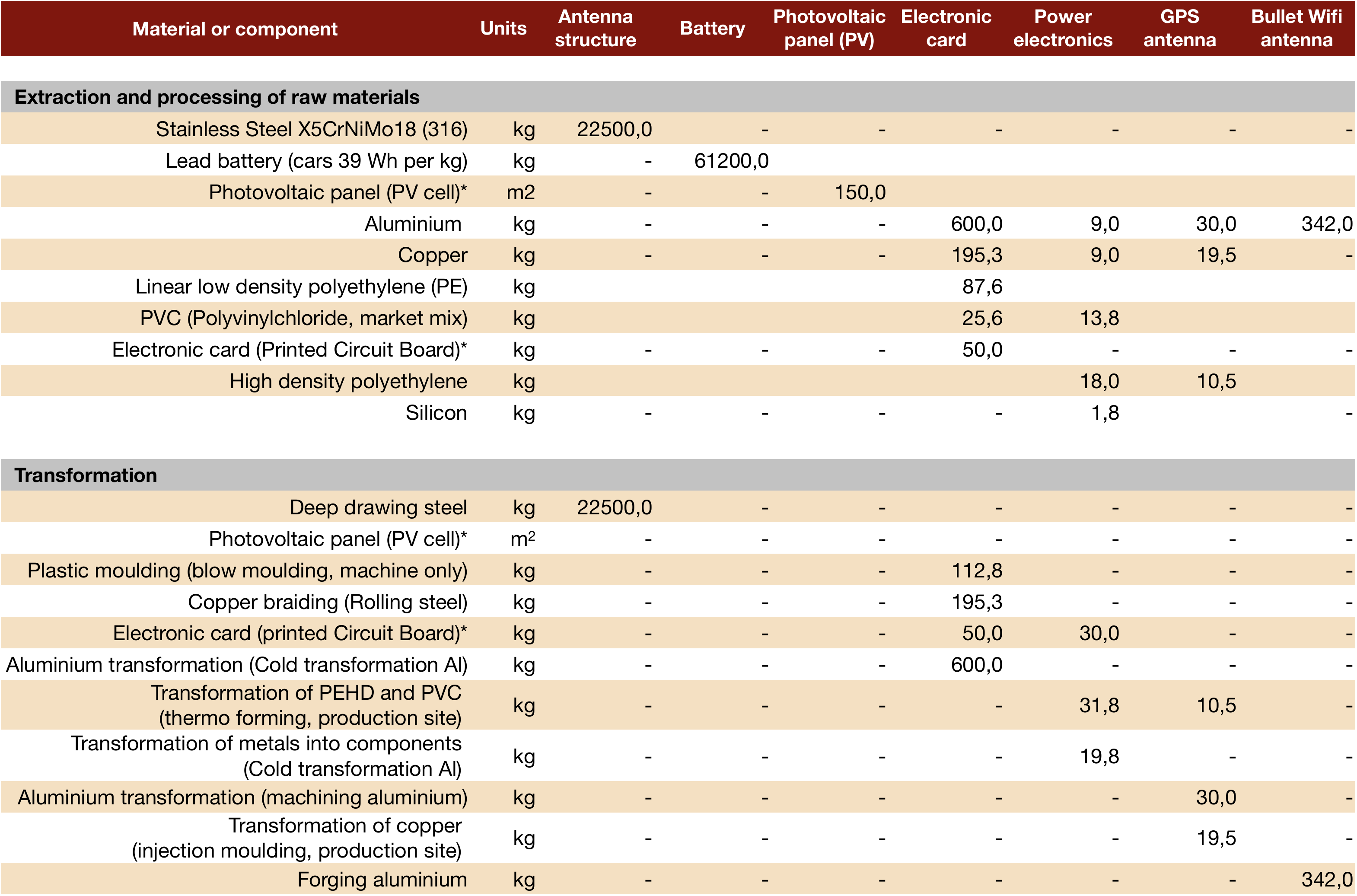}
    \caption{Main inventory inputs for the GRANDProto300 detection units. The assumptions on components are detailed in Section~\ref{section:assumption_components}. *Some activities/processes in the Idemat database (PV cell, and electronic card) include upstream activities (i.e. also the raw materials) in the activity impact factor. Since the percentage of the environmental impact corresponding to extraction/transformation is not detailed, in this work 50\% of the impact is assigned to each (extraction and transformation).}
    \label{table:inventory_main}
\end{table*}

\begin{table*}[!tb]
    \centering
    \includegraphics[width=0.95\textwidth]{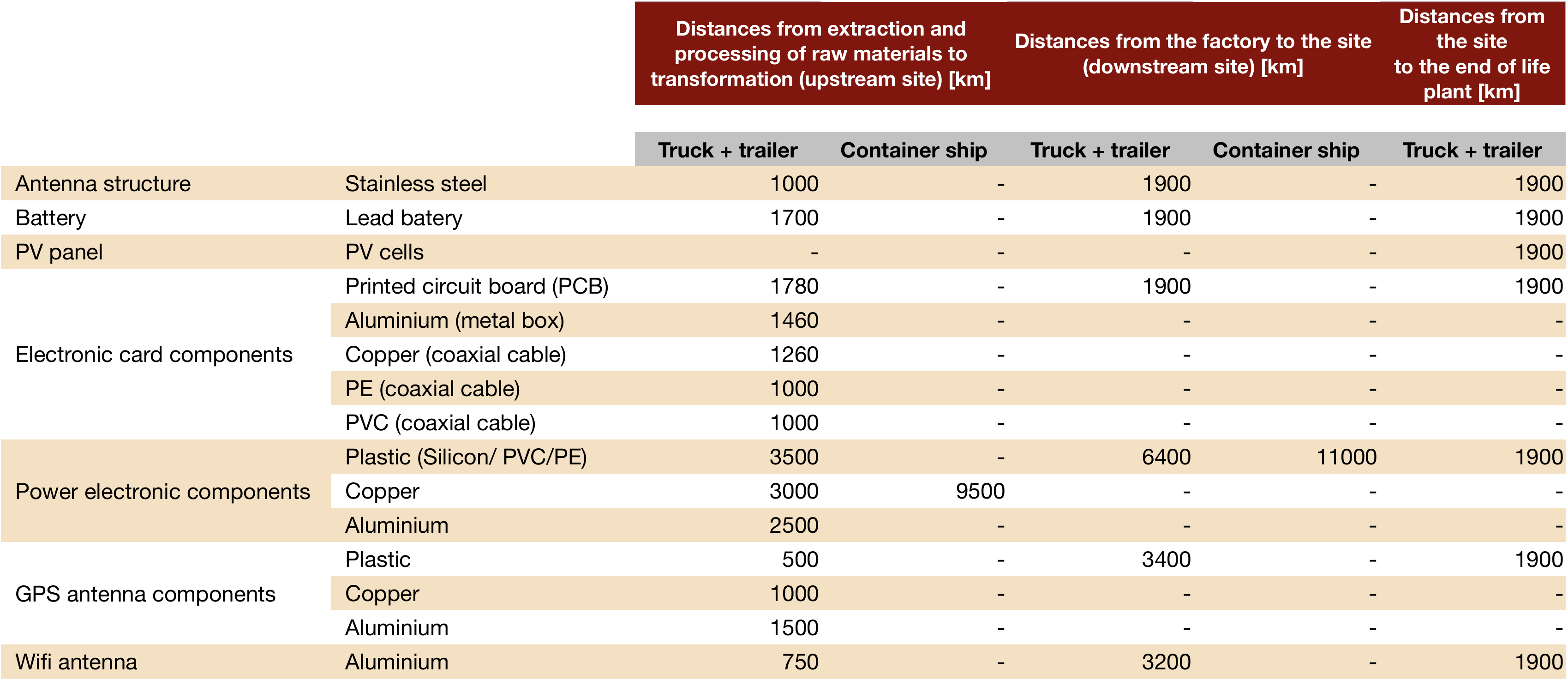}
    \caption{Transportation modes and distances for the GRANDProto300 components. Assumptions are detailed in Sections~\ref{section:assumptions_transport} and \ref{section:assumption_components}. 
    }
    \label{table:inventory_transport}
\end{table*}

\subsubsection{Antenna Structure}
The antenna is made of 120\,kg of stainless steel. 
To have a general perspective, and for comparison purposes, we use the data of the most common stainless steel for outside use: stainless steel 316. Note that for the production of a first batch of 100 GRANDProto300 antennas, another type of stainless steel X20Cr13 (420) was used, which improves the environmental impact, as discussed in Section~\ref{section:action_plans}.

For upstream transportation: China being the first steel producer in the world, and Hebei province the major producer in China, we assume the steel to be produced in that region. The transformation of stainless steel into an antenna is performed in Xi'An, following a `deep drawing' process, which enables to model steel into a given shape. The average distance from the Heibei province to the Xi'An factory is 1000\,km.

For the end-of-life, it is assumed in this study that stainless steel is collected and recycled. Environmental impact coefficients for the collecting and recycling are available in the Idemat database. The non-recyclable part of the stainless steel (56\%) is assumed to be treated by conventional end-of-life methods in China ; the impact factor for modeling the waste scenario is taken from Ecoinvent \cite{Ecoinvent2021}.

\subsubsection{Battery}
A lead-acid battery will store the electricity produced by the PV panel. The battery is composed of one positive plate in lead dioxide and one negative plate in lead only. Each battery weighs a total of 51\,kg. 

Our calculations indicate that the battery has an energy density of 36.3 Wh/kg approximately. For our estimates, we use the emission coefficients of the lead battery of a car 
with similar energy densities (39\,Wh/kg), which are provided in the Idemat database. This Idemat activity only takes into account the materials required to produce the battery; due to the lack of information, the transformation process of the battery is not considered. 

Most of the lead battery factories in China are located in the region of Guangdong, close to Hong-Kong. Therefore, we assume for upstream transportation that the batteries are manufactured in this area, 1700\,km away from Xi'An.

One important factor to take into account is battery life, which is estimated by the manufacturer to be around 5 years for field use. For an experiment duration of 20\,years, this implies the use of 4 batteries per antenna. The impact of the battery production and transportation is hence multiplied by 4 in our estimates, which contributes to significantly increase the global impact.

At the end-of-life, batteries can be recycled. In case the battery is recycled, we assume that 75\% of the lead of the battery can be collected, sorted out and recycled. Given the lack of information about the recycling process, we assume the environmental impact factor provided for lead recycling in the Idemat database. For the end-of-life of the remaining materials an approximate impact factor is considered \cite{Ecoinvent2021}.

\subsubsection{Photovoltaic (PV) panel} \label{section:PV_panel}

For the impacts of the photovoltaic panel, we consider that 300\,m$^2$ of PV are required. The materials extraction and transformation are accounted in the coefficient based on the Idemat database (Idematapp2021 PV cell). Each poly-crystalline solar module has an approximate area of 1\,m$^2$ and a weight of 10 kg.  

For the end-of-life of the solar panels, it is considered that each module is mainly made of glass (approx. 70\%) and aluminium for the frame (approx. 18\%). The materials and their respective quantities are calculated from the \cite{Latunussa2016} report. The glass, aluminium, copper, polymers, silver and various metals are collected, sorted and recycled. The environmental impact coefficients available in the Idemat database are used. The environmental impacts for glass recycling is not available in the Idemat database. For glass and for materials with no disposal method available in the database, a conventional disposal method is assumed \cite{Ecoinvent2021}.

\subsubsection{Electronic Card}\label{section:electronic_card}
The electronic card comprises the card itself, as well as the metal box in which it is contained, and the coaxial cables linked to the card.

\begin{description}

\item[\bf Electronic card]

The electronic card consists of 326\,g of {\sc Megtron6}\footnote{\href{https://www.matrixelectronics.com/products/panasonic/megtron-6/}{www.matrixelectronics.com/products/panasonic/megtron-6/}}, a  material made of copper, glass fiber and resin. The details of the production processes of the card are not available. Hence we assume environmental factors for an average PCB (Printed Circuit Board) available in the Idemat database. This activity/process (Printed Circuit Board - electronic card) includes the raw materials extraction and transformation. The assembly and testing of the electronic card are not considered in this work.

We calculate the residual upstream transportation impacts based on our knowledge that the PCB is produced by Shenzhen Suntak Multilayer PCB Co., Ltd. a manufacturer based in Shenzhen, Guangdong, 1780\,km away from the Xi'An factory.

Concerning the end-of-life, the collection of the different materials that compose the electronic card (copper and glass fiber) is considered, as well as the recycling of 44\% of copper. The end-of-life of plastics is accounted for as upcycling. The emission factors are extracted from the Idemat database. Due to lack of information, the disposal of glass fiber is included using an approximate emission factor, as well as the non-recyclable part of the copper \cite{Ecoinvent2021}.

Concerning the end-of-life, glass fiber and copper must be collected, sorted and recycled. 

\vspace{0.2cm}

\item[\bf Metal Box]

The metal box is produced by Ningbo Uworthy Electronic Technology, a professional case enclosure manufacturer. We consider that the box is entirely made of aluminium and weighs 2\,kg.

For upstream transportation, it is assumed that the aluminium stems from the major extraction site in China, located at Jinan, 825\,km away from the factory in Cixi (Ningbo). The distance between the two factories of Cixi and Xi’An is 1460\,km.

Concerning the transformation process, environmental impact coefficients of machining aluminium are used.

Finally, the collection and recycling (34\%) of aluminium are considered at the end of the project using coefficients available in the Idemat and Ecoinvent databases.
\vspace{0.2cm}

\item[\bf Coaxial Cables]

Three cables (4\,m long each) connect the electronic card to the antenna, one cable of 2.5\,m long connects the Wi-Fi antenna to the electronic card, and another cable connects the GPS antenna to the electronic card. These cables are composed of four layers: copper, polyethylene (PE), copper again and poly vinyl chloride (PVC). After calculations based on the surface of each layer, we can approximate the total mass of the materials which compose the 5 cables: 651\,g of copper, 292\,g of polyethylene and 84\,g of PVC.

For upstream transportation impacts, it is assumed that copper is produced on the site of Dexing (East of China) since it is one of the major production plants in China, located at 1260\,km from the factory in Xi'An. Polyethylene and PVC are produced everywhere in the country, thus we make the assumption that a potential producer is 1000\,km away from Xi'An.
Coefficients for transformation processes concerning copper (braiding) and plastics (moulding) are available in the Idemat database.

For the end-of-life stage, we assume that the cables are all collected and recycled. Environmental impact coefficients for the upcycling and recycling of plastic (PE and PVC) and copper, respectively, are also available in the Idemat database.  The impact factor considered for the non-recyclable part of the copper is extracted from the Ecoinvent database (approximate value of waste management in general) \cite{Ecoinvent2021}. 

\end{description}

\subsubsection{Power electronics}

The power electronics corresponds to the components needed to connect the PV panel to the battery. The component implemented for this function is Genasun GV10-MPPT. It is composed of a plastic box encasing the controller system which includes a PCB and electronic devices. It weighs 186\,g and is assembled in Arlington, WA, USA.  

From its dimensions, we estimate the mass of the plastic box to 30\,g and assume that it is composed of High Density Polyethylene (HDPE), the most commonly used thermoplastic. The manufacturer neither provides a list of the electronic components nor of their materials. We assume the following composition for the electronics: 15\,g of copper, 15\,g of aluminium, 3\,g of silicon and 23\,g of plastic (protection of components), as well as approximately 100\,g of PCB.

Chile is one of the main producers of copper, and the transportation selected from Chile to the USA is a container ship over 9500\,km, then 3000\,km by 24-ton truck. Canada is the closest main producer of aluminium, with transportation over 2500\,km by 24-ton truck to the USA.
All other materials and PCB are assumed to be produced in the USA, so that transportation by truck is considered.

For plastic materials, we assume that they are thermoformed. For metals, because of the scarcity of documented processes in the database, we choose to apply an aluminium transformation process to all metals. Concerning the PCB, the transformation is already taken into account in its coefficient.

The power electronics is transported by container ship from the USA to China over 11000 km, and then by truck to Xi'An over 1500 km from the port to Xi'An.

For the end-of-life, we consider that the Genasun GV10-MPPT is collected, copper and aluminium are recycled. For the plastics (PVC and PE) the upcycling process from Idemat database is used, while for the non-recyclable materials and silicon, a end-of-life coefficient is taken from Ecoinvent database \cite{Ecoinvent2021}.

\subsubsection{GPS antenna}
The GPS antenna enables to record the precise time and location of received electromagnetic signals. The model implemented in the prototype units is the Trimble Bullet III GPS antenna. 

No detailed information is available on the composition of the GPS antenna. From the approximate dimensions of the device and its typical engineering design, we estimate that the GPS antenna is composed of approximately 35\,g of HDPE for the thermoplastic enclosure, 100\,g of aluminium and 65\,g of copper.

For upstream transportation, we consider that plastic can be produced anywhere in China and should be transported over less than 500\,km. Copper is extracted in China (second world producer after Chile). We consider that copper comes from the province of Jiangxi in China and is transported over less than 1000\,km. Aluminium can be produced in China, so near the GPS antenna factory. 
The location of the GPS antenna factory is uncertain, hence we assume that the GPS antennas are produced in Jiangsu (about 1500\,km from Xi'An), where one of the factories of the manufacturer is located, as well as several other aluminium companies. 

For the end-of-life stage, we assume that all the materials are recycled and that the location of the recycling site is near Xi'An.

\subsubsection{Bullet Wi-Fi antenna}\label{section:wifi_antenna}

The Bullet Wi-Fi antenna is mainly made of an antenna reflector grid, a feedhorn assembled to the front thanks to an aluminium L bracket, some screws and washers. Because of the lack of information on the other lighter components, we choose to approximate the overall Wi-Fi antenna (of weight 1.14\,kg) by the reflector grid, considering that impacts of the assembly stage are negligible. 

According to the company in charge of the assembly of the antenna (L-Com) the aluminium reflector is die-casted. The Idemat database does not provide environmental impact information about this process, therefore we choose the closest documented process : forging.

For the upstream transportation, it is assumed that the factory in charge of the die-casting is located at Jinan (Shandong), the biggest aluminium production area in China and also the closest of the L-Com factory located in Suzhou, where the product is assembled. The distance between Jinan and Suzhou is 750\,km, and of Suzhou to Xi'An 1300\,km. For the end-of-life stage, we consider that the primary aluminium is recycled. 

\section{Results and discussion}\label{section:results}

 This section presents and discusses the findings of the environmental impact assessment for the different stages (raw materials extraction, transformation, transport and end-of-life) of the 7 main components of the GRAND prototype detection units. In this work, the ``operation'' stage (use phase, downstream) is considered to have a non-significant environmental impact, since the detection units operate autonomously with solar and battery power. 
 
 As mentioned above, the Idemat method is used in this work to perform a simplified LCA. In order to illustrate the relative impacts for the different components, the results are normalized. The normalization is performed by dividing each of the results for the different impact categories into the respective normalization factors. It allows us to compare the midpoint characterization results since the normalized results share the same units,  in order to select the most relevant categories and analyze them in detail (mid-point indicators, \ref{section:impact}). In this study the factors proposed by the Joint Research Centre Technical Report of the European commission are used \cite{Sala2018}. Table~\ref{table:impacts} shows the different values.
 
 \subsection{Normalized results for the components of the GRANDProto300 detection units}
 
\begin{figure*}[tb]
    \centering
    \includegraphics[width=0.8\textwidth]{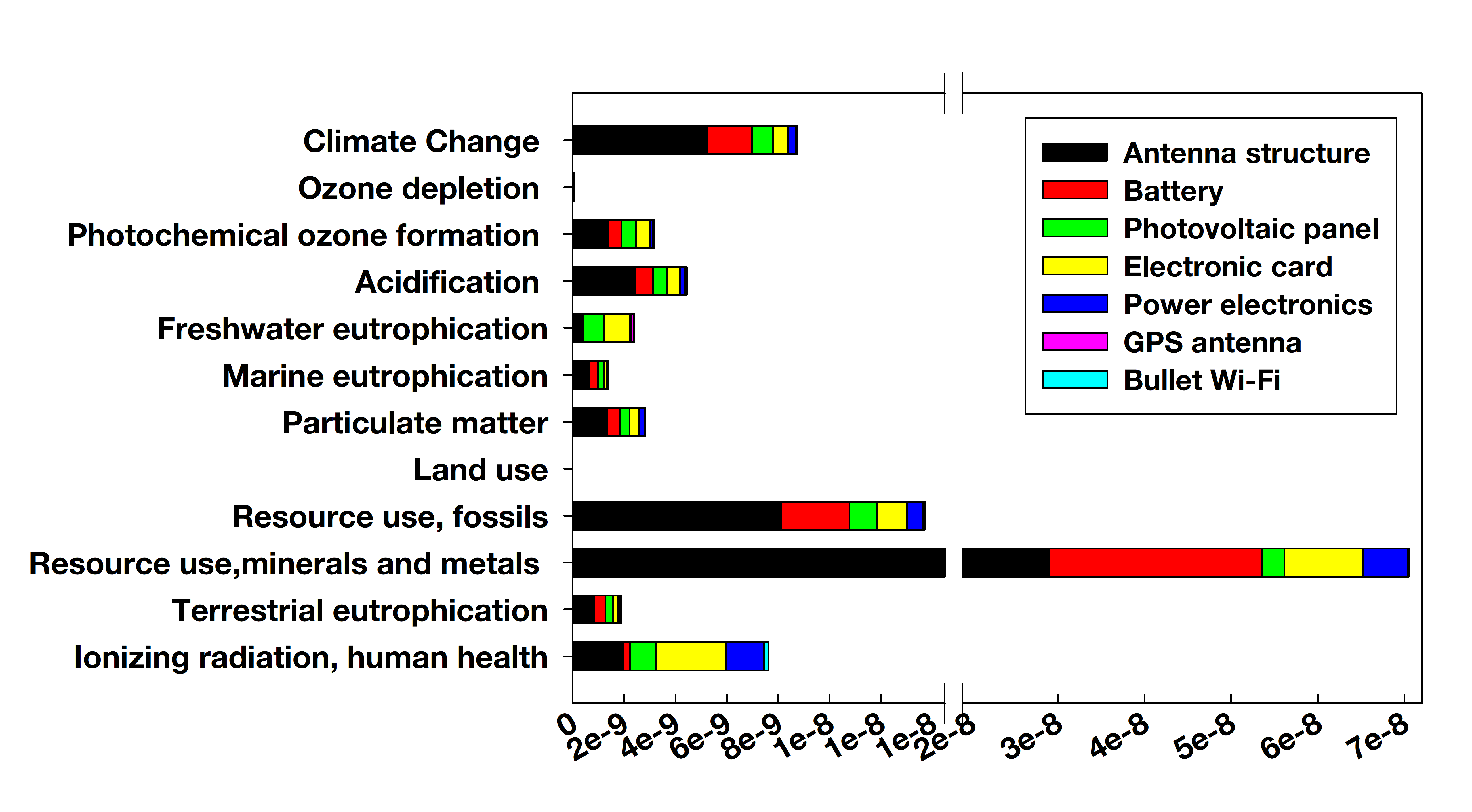}
    \caption{Normalized impacts in different midpoint categories for GRANDProto300.}
    \label{fig:normalized}
\end{figure*}

The assessment result for GRANDProto300 illustrated in Figure~\ref{fig:normalized} shows that the components with the highest environmental impact are the antenna structure and the battery 
with contributions distributed differently for each one of the categories. 
 
Based on the normalized results it is possible to identify the most pregnant indicators, which in this study are ``resource use, minerals and metals''; ``resource use, fossils''; ``ionizing radiation, human health''; ``climate change''; and ``acidification''. Although the other impact categories are not insignificant, this study emphasizes the five identified. 
 
 \begin{figure*}[!tb]
    \centering
\includegraphics[width=0.8\textwidth]{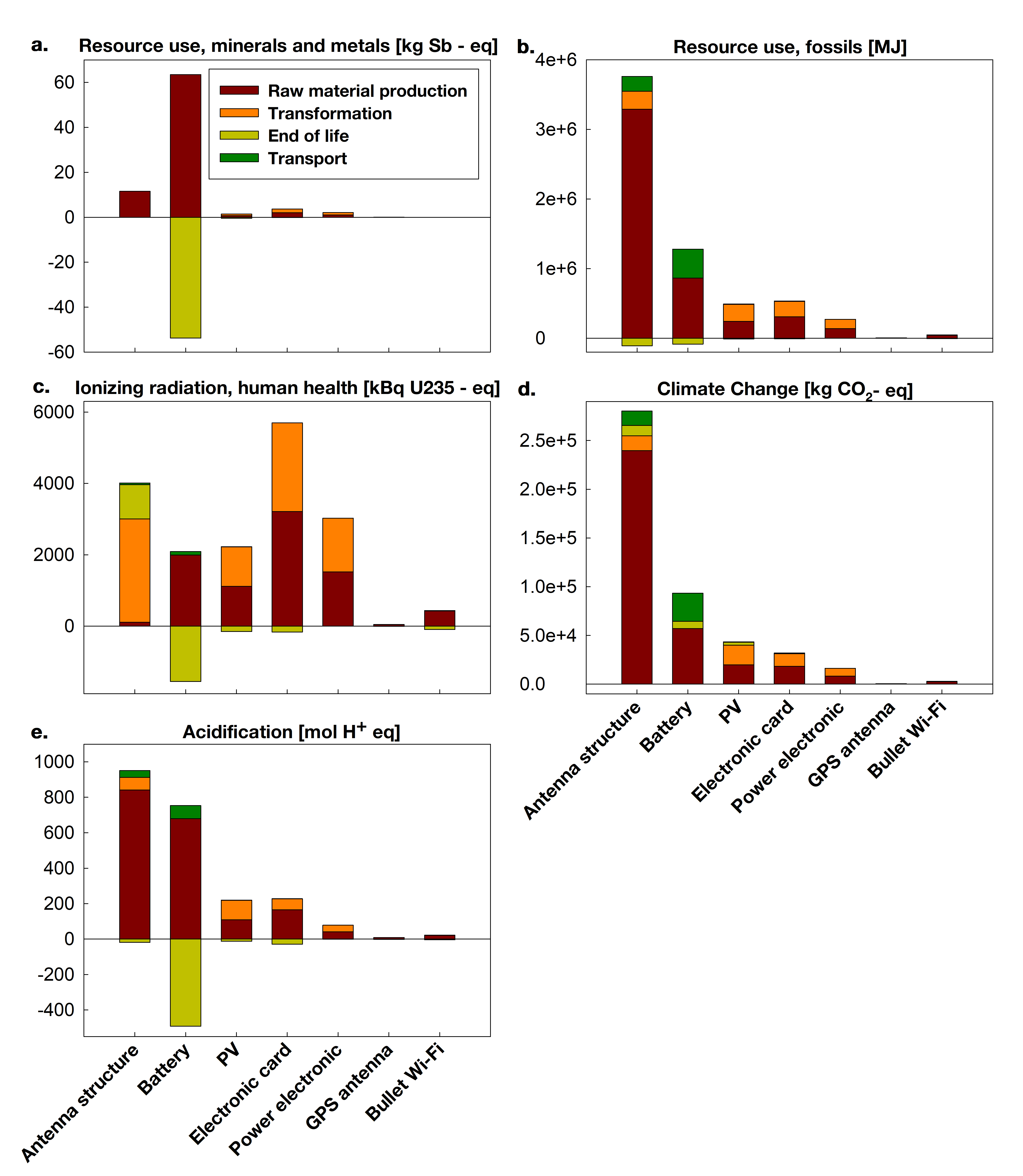}
    \caption{Impact level (with indicated units) for components and life cycle processes of GRANDProto300, in relation to the LCA categories. The end-of-life stage is represented with a negative impact value.}
    \label{fig:5indicators}
\end{figure*}

\subsection{Impact assessment results: different components  and processes of the GRANDProto300 detection units} \label{section:impact}

``Resource use, minerals and metals'' is the most impacting category, according to the normalized results. Figure~\ref{fig:5indicators}.a shows that the raw material extraction is the process with the highest contribution to this indicator. For the battery, which is the component with the greatest contribution, a particular peak is present in the figure due to the extraction of materials (material transformation is not considered in the case of the battery as no information is available). Furthermore, it is observed that by recycling 75\% of the lead in the battery, the environmental impact on the resource use, minerals and metals indicator is reduced to approximately 10 kg SB-eq. It is also possible to observe that the extraction of materials (stainless steel) for the construction of the antenna structure represents an important contribution to this indicator.

For the categories ``resource use, fossils'', and ``climate change'', figure~\ref{fig:5indicators}.b and d, the raw material production is the process with the greatest contribution. In these two categories, the component with the highest environmental impact is the antenna structure due to the production of stainless steel. In the case of the battery, the environmental impacts are due to the raw materials extraction and transport, 61\% and 31\%, respectively. Transport has a significant contribution in these categories because of the lifetime and the amount of batteries.

Our results for the climate change category can be compared with the carbon footprint results obtained by Aujoux et al.\cite{Aujoux2021}. The authors evaluated the carbon footprint of the materials used for the antenna structure, solar panels and battery. They found that the impact associated with the production of the antenna structure for GRANDProto300 project is 65\,tCO2-eq for the 300 units. While in this work the steel of the antenna structure has a contribution of approximately 240\,tCO2-eq. This difference is due to the fact that Aujoux et al. use an average emission factor of 2.9 kg CO2-eq/kg stainless steel \cite{Aujoux2021}, while in this work 316 stainless steel was selected as the material with a factor of approximately 6.7 kg CO2-eq/kg stainless steel. Additionally, this work considers that the structure has a weight of 120\,kg, instead of 75\,kg in the other study \cite{Aujoux2021}. The assumptions of the two studies differ, as they were revisited for the LCA after conception changes in GRANDProto300 in-between.

In terms of the materials needed to produce the battery, the results of the two studies are equivalent, 45.9 kg CO2-eq/battery. However the authors consider that 300 batteries are used for the GRAND-Proto300 and do not consider the lifetime of the batteries. As previously mentioned, in this present work it is considered that 1200 batteries will be used for the whole array.

The raw materials production and transformation processes are also the ones that have the major contribution in the ``ionizing radiation, human health category'', with a distribution among the different components of the GRANDProto300 detection units, as is shown in figure~\ref{fig:5indicators}.c. In the case of the antenna structure, the main environmental impact is associated with the transformation and the stainless steel recycling process. For the photovoltaic panels, the production of materials, as well as the transformation of PVs have the greatest environmental impact. The same impact is assumed for the processing and extraction of materials. The component with the greatest impact in this category is the electronic card, due to the production of materials and transformation of the Printed Circuit Board (PCB). PCB production and transformation are also the processes responsible for the impact associated with the power electronics component.  

Finally, figure~\ref{fig:5indicators}.e shows that for ``acidification'', the main impacts are associated with the production of materials. The greatest environmental impact in this category is associated with the production of the raw materials for the antenna structure. Likewise, the production of raw materials for the battery have a considerable environmental impact, due to the use of lead batteries.

\subsection{Limitations and discussions}\label{section:discussions}

The major limitation of our analysis stems from the lack of detailed information on the materials used by the suppliers (for example, metal alloys), or on specific quantities of raw materials used to manufacture components such as power electronics, GPS or Wi-Fi antennas. This applies to all components, and varies for the different manufacturers. Therefore, in this research, assumptions have been made about the materials and quantities used. 

The metal composition is also difficult to define as the mix between primary metal (from raw materials) and secondary metal (from recycled materials) varies from a country to another and from an industry to another. In order to maintain consistency with the different components, we always consider the use of primary materials. 

Finally, for raw materials, the extraction site can be questioned. Manufacturers near the coast can be supplied with foreign metals transported by ship, which would lead to differences in transportation impacts. We have not made this assumption as China has huge extractive deposits and relatively low production costs. The question of the origin of raw materials will be all the more important in the future stages of the project, especially with GRAND200k, since the location of production sites may vary according to the location of the GRAND sub-arrays on the Earth.

Our choices in terms of transportation can also be questioned. For this preliminary study, we choose to use only road transportation. In reality a case-by-case study of the transportation method used should be done for each component. Indeed, the volume of the component, the stock capacity of manufacturers, the proximity with railroads or the topography of the area influence the transportation method selected by manufacturers and transporters. 

Concerning the recycling phase, we assume different recycling percentages for each material, considering the information available in the Idemat database, which reflects the recycling processes currently available. 
It is likely that the percentages of recycled materials will change over the decades due to potential new regulations and new technological processes. It is worth mentioning that there is no available information in the Idemat database regarding the recycling of polluting components such as electronic cards or solar panels. The limitations of the analysis of the recycling process also relate to the lack of available information on the final disposal processes of some components and the end-of-life management of the GRAND project.

Lastly, the environmental impact or contribution of the end-of-life for some materials was evaluated in a simplified way  due to the lack of current available information.

\section{Hotspots identification and potential action plans}\label{section:action_plans}

The environmental hotspots can be identified after the LCI and the simplified LCIA. As is shown in Figure~\ref{fig:5indicators}, the raw materials for the antenna structure, battery and PV, as well as the transformation process of these components represent the main opportunity areas to propose alternatives that allow a reduction in the environmental impact of the future stages of the GRAND project, based on the GRANDProto300 phase. 

It is important to note that impact reduction actions generally require to make a trade-off between various environmental impacts. This means that it is necessary to first select the indicators with the greatest environmental impacts (Section~\ref{section:impact}), and then make sure that, with the different potential action plans, there is no considerable displacement of environmental burdens from one category to another. 

\subsection{Changing the antenna structure}
The first potential action to consider relates to the antenna structure. The antenna itself (considering its entire life cycle) represents approximately 60\% of the ``climate change'', and ``resource use (fossils)'' categories, while for the ``ionizing radiation (human health)'' and ``acidification'' categories, it represents 25\%  and 55\%, respectively, of the total impact. More specifically, the extraction phase represents more than 90\% of the environmental impact in almost all the categories evaluated, except for ``ionizing radiation, human health''. 

\subsubsection{Limiting the mass of the antenna structure}
The first solution therefore is to limit the mass of the raw material used. For example, a reduction of 6 kg per antenna (using 114 kg antennas) would reduce the environmental impact by 6\% in all the categories evaluated.  But by limiting the mass, the strength or the length will be affected. This choice is crucial for the success of the project and therefore must be studied by researchers in order to evaluate the impact on data acquisition of a shorter or less stable structure. 

\subsubsection{Changing the alloy of the antenna structure}
In terms of the antenna structure, another alternative to reduce the environmental impact associated with the extraction of its raw material is to change the metal (alloy). A sensitivity analysis is performed to evaluate different potential alloys. The results are presented in Table~\ref{table:sens_analysis}, in terms of Average Absolute Percentage Deviation (AAPD) for each alloy and each impact, estimated as
\begin{equation}
{\rm AAPD} \,[\%] = \frac{\mbox{Value base case} - \mbox{ Value alloy}\,i}{\mbox{Value base case}}\times 100\ .
\end{equation}
Table \ref{table:sens_analysis} shows that it is possible to suggest the use of two different stainless steel alloys, reduce the total environmental impact of the GRAND project between 5\% and 34\%  approximately in the climate change and resource use, fossil, while in the resource use -minerals and metals-- and the acidification indicators the reduction is 83\% and 18\%, respectively for the alloy 1: X5CrNi18 (304), and 66\% and 22\%, respectively for the alloy 2:  X20Cr13 (420). 
Therefore, for future phases of the project it is suggested to using the alloy X20Cr13 (420).

\begin{table}[tb]
    \centering
    \includegraphics[width=\columnwidth]{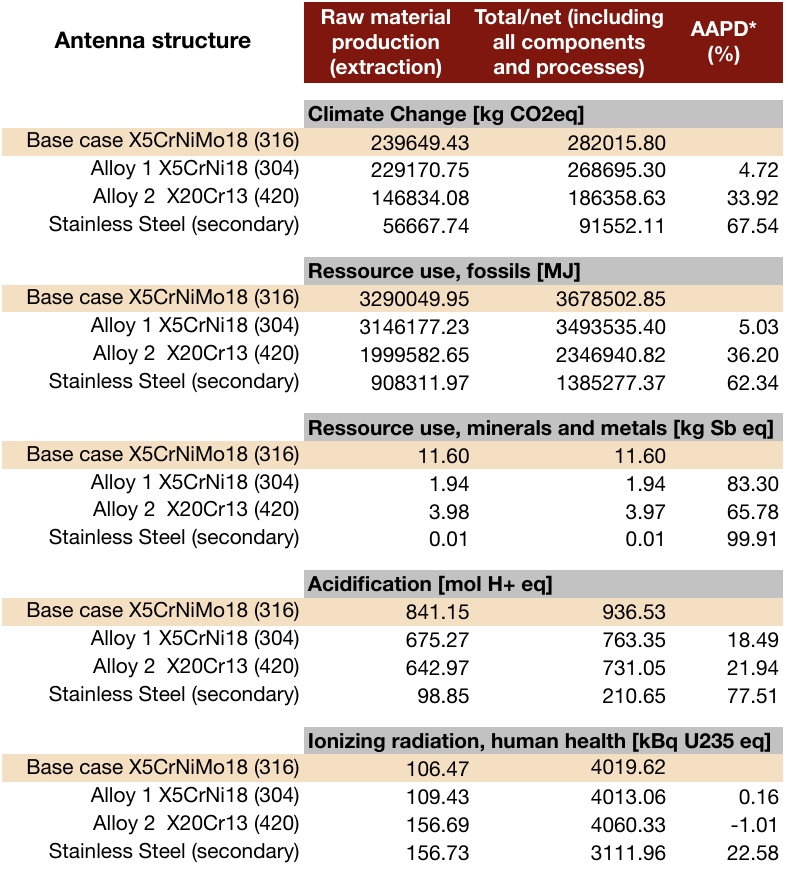}
    \caption{Influence of using different stainless steel alloys for the antenna structure on the environmental impact. *AAPD [\%]: Average absolute percentage deviation.}
    \label{table:sens_analysis}
\end{table}

\begin{table*}[!tb]
    \centering
    \includegraphics[width=0.95\textwidth]{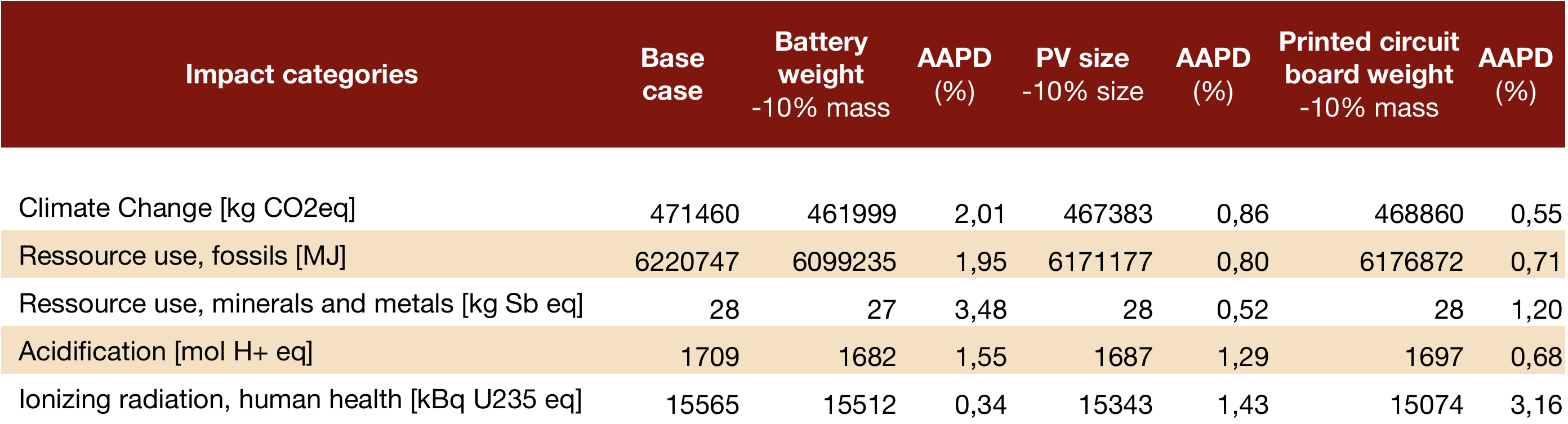}
    \caption{Parameter influence (battery and PCB weight, PV size) on the environmental impact of GRANDProto300.
    }
    \label{table:sens_analysis2}
\end{table*}

\subsubsection{Use of secondary stainless steel (recycled)}\label{section:use_sec_stainless}
Another alternative to reduce the environmental impact of the antenna structure is the use of recycled stainless steel. Table \ref{table:sens_analysis} shows the difference between the different impact categories evaluated when using primary and secondary materials.It is possible to observe that there is a reduction between 22\% and 99\% in the different indicators. Therefore, based on this analysis it is possible to suggest that the most viable option in terms of environmental impact reduction is the use of recycled materials, all the more as it avoids any change in the antenna structure. 

\subsection{Changes concerning the battery}

\subsubsection{Improving the battery life time}\label{section:improve_battery}
Additionally, it can be interesting to work on the battery. Indeed, according to the technical document of the GRAND project, the batteries have a useful life of five years, although it depends on the number of charge-discharge cycles, the initial state of the battery, the usage conditions and the technology. Thus, four different batteries per antenna will be needed during the 20 years of the GRAND project. An increase in battery life would entail a reduction in the environmental impact associated to the use of batteries, and it would reduce the number of journeys to the site of the antennas, thus reducing the transportation distances. This improvement would be the greatest for the GRAND200k phase, since the distances will be longer: reducing maintenance to a minimum appears to be essential.

\subsubsection{Use of remanufactured batteries}

In this work it is not possible to evaluate the environmental impact of the GRANDProto300 project when using remanufactured batteries, due to the lack of available information. However, based on the work done by Kim et al. \cite{Kim2019} it is possible to suggest that the use of remanufactured batteries is feasible and allows reducing the impact by up to 40\% for different categories. As in \ref{section:use_sec_stainless}), this again suggests that one of the main improvement options for future phases of the project is the use of secondary or remanufactured parts and materials.

\subsection{Recycling PV panel components at the end-of-life}
As for photovoltaic panels, potential improvements may stem from the recycling processes of glass and silicon, the main components of PVs. It has been observed that recycling reduces the environmental impact of the use of photovoltaic panels, because the production of some raw materials such as aluminum, raw materials for the production of primary white glass, copper, silicon or silver is avoided, thanks to the recycled material from the treatment of photovoltaic waste  \cite{Latunussa2016}. 

\subsection{Evaluate manufacturing options for the Print Circuit Board}
Regarding the components with less environmental impact, i.e. the electronic card and the power electronics, the critical part is the use of the Printed Circuit Board. It can be suggested to evaluate different manufacturing options, and evaluate the processes available for the end-of-life of the PCB. This area of improvement is supported by the fact that some authors, such as Ref.~\cite{Nassajfar2021}, have suggested that the use of different materials can reduce the environmental impact in the categories of climate change and resource use- fossils by up to 50\%. The end-of-life of the PCB is not evaluated in this paper because no information is available in the Idemat database.  

\subsection{Improving the supply chain}
Also, an improvement in the supply chain can be proposed. It can be reached through a reduction in transportation distances for example, as it is found that for GRANDProto300 transportation contributes about 12\% to the categories of climate change and resource use (fossils).
According to a sensitivity analysis, concerning the transport, it is possible to suggest that reducing the distance until the waste treatment plant by approximately 30\% would reduce the total environmental impact by approximately 3\% in the climate change and resource use (fossils) categories. This optimization of distances is important, not only for the end-of-life stage, as longer transport distances may be contemplated in GRAND200k.

\subsection{Recycling or remanufacturing materials}
Another improvement area of the process is related to the end-of-life of materials. It is important to consider recycling or remanufacturing processes at the end-of-life of the different stages of the GRAND Project, since a proper disposal of the different materials and components of the project at the end of their life could potentially reduce the environmental impact of the project. However, due to the lack of available information as mentioned in Section~\ref{section:discussions}, it would be necessary to collect experimental data to evaluate the real environmental impact related to the recycling or remanufacturing processes.  

\subsection{Use secondary or recycled materials}
One of the most feasible alternatives to reduce the environmental impact of the GRAND project is to use secondary and recycled materials for the production of the different parts and components. Indeed, as previously mentioned in Sections~\ref{section:use_sec_stainless} and \ref{section:improve_battery} there could be a significant reduction in the different indicators by using recycled or remanufactured raw materials. In this work this alternative is only presented for the materials that compose the antenna structure and the battery. However, based on the results, this alternative could be considered for other materials/parts of the future stages of the project. 

\subsection{Resizing some components}
Finally, for future stages of the project it could be suggested to redesign some components in order to reduce the use of materials for their manufacture.  However given the lack of information on this matter, a sensitivity analysis is performed to evaluate the influence of this proposal with respect to the results of the life cycle analysis. Table \ref{table:sens_analysis2}, presents the results for the different indicators, as well as the AAPD to compare with respect to the base case. From this "one-at-a-time" analysis it is possible to observe that with a 10\% reduction in the size/mass of the battery or PV or antenna, the difference in the result is not significant for most of the cases. However, it is important to note that the size of all components is related (i.e. if the size of the panel changes, the size of the battery changes), so it could be suggested that by reducing the size/mass of different components by 10\%, the environmental impact of the project could be reduced by about 5\%. It is suggested that the feasibility of this improvement option be studied for future phases of the GRAND project, due to the scale of the future phases.

\section{Conclusions and perspectives}\label{section:conclusions} 

This study brings first results about the environmental impact of a scientific project at its pilot scale. Indeed, the LCA of the GRANDProto300 project shows that the main environmental impacts relate to resource use, minerals and metals, and fossils, as well as climate change. They primarily stem from the raw material production and the transformation of the antenna structure and the battery. The results suggest action plans to reduce these impacts, e.g. change the weight/size for the structure of the antenna, use a different alloy, and use recycled materials/components. 

Most of these actions would benefit the experiment from other points of views (robustness, costs), and hence are in line with the goals of the GRAND Collaboration: for instance, reducing the size of the detection units will lead to a more reliable setup, as smaller sizes indeed limit the mechanical burden, hence reduces failure chances. 

The work already achieved can be continued and become even more precise and exhaustive. Firstly, more precise information concerning the origin of the components can improve the reliability and the accuracy of the study. In particular, the origin of the raw materials, the location of the manufacturers and the location of the antennas (for GRAND 200k) can improve our results. Scientists taking part in the project will be able to provide this information and complete our work in the following years when all the components are produced at an industrial scale.

Another aspect that can be improved relates to the database. In this study we use Idemat, a free and simplified database. For this reason, the assumptions made in this work are based on the information available in Idemat. It is important to note that the results of this work have been compared with those of SimaPro using the Ecoinvent 3.8 databases. The tendency of the results is similar.

Finally, it would be interesting in future studies of the different stages of the project, to include installation and maintenance in the system boundaries as well as the use of digital tools, because these processes will become increasingly important as the project expands.

The work that we have achieved for the GRAND project is new in the research community. The environmental impact of research projects is marginally studied. The initiative started with the GRAND project may pave the way to more environmental awareness and footprint reduction when carrying out research projects.

\section*{Acknowledgements}
The authors thank the GRAND Collaboration for support on this project, in particular Pengfei Zhang and Charles Timmermans for valuable input on the detection unit design and production, and the whole. The authors are also grateful to the referee who contributed to the improvement of the study through very useful comments.
Kumiko Kotera was supported by the APACHE grant (ANR-16-CE31-0001) of the French Agence Nationale de la Recherche and the CNRS Programme Blanc MITI (2023.1 268448 GRAND; France).

\appendix
\section{Units of environmental impacts}\label{app:units}

\begin{description}
\item[\bf kg CO2 eq] Mass of Carbon dioxide equivalent. Greenhouse gases that highly contribute to global warming. All these gases are converted into carbon dioxide quantities using their global warming potential. 

\item[\bf kg CFC-11 eq] Mass of trichlorofluoromethane equivalent. Several greenhouse gases contribute to the destruction of the ozone layer (CFC, HCFC, Halons…). They are converted into a quantity of trichlorofluoromethane using their ozone depletion potential.

\item[\bf kg NMVOC eq] Equivalent mass of non-methane volatile organic compounds. These are a large variety of chemically different compounds, such as benzene, ethanol or acetone.

\item[\bf mol H+ eq] Amount of equivalent hydrogen proton available in a molecule. This quantity is obtained from the acidification potential.

\item[\bf mol N eq] Amount of equivalent nitrogen available in a molecule. 

\item[\bf kg P eq] Equivalent mass of phosphorus.
\item[\bf kg Sb eq] Equivalent mass of antimony.

\item[\bf Pt] Point (kg C deficit). 
\item[\bf m3 water] Water volume.
\item[\bf MJ] Energy in mega joule.
\item[\bf kBq $^{235}$U eq] Radioactivity in units of $^{235}$U radioactivity.
\end{description}

\bibliography{biblio} 

\begin{thebibliography}{10}
\expandafter\ifx\csname url\endcsname\relax
  \def\url#1{\texttt{#1}}\fi
\expandafter\ifx\csname urlprefix\endcsname\relax\def\urlprefix{URL }\fi
\expandafter\ifx\csname href\endcsname\relax
  \def\href#1#2{#2} \def\path#1{#1}\fi

\bibitem{Mariette2022}
J.~Mariette, O.~Blanchard, O.~Bern{\'e}, O.~Aumont, J.~Carrey, A.~L. Ligozat,
  E.~Lellouch, P.-e. Roche, G.~Guennebaud, J.~Thanwerdas, P.~Bardou, G.~Salin,
  E.~Maigne, S.~Servan, T.~Ben-Ari,
  \href{https://iopscience.iop.org/article/10.1088/2634-4505/ac84a4}{An
  open-source tool to assess the carbon footprint of research}, {Environ. Res.:
  Infrastruct. Sustain.} 2 (2022) 035008.
\newblock \href {http://dx.doi.org/10.1101/2021.01.14.426384}
  {\path{doi:10.1101/2021.01.14.426384}}.
\newline\urlprefix\url{https://iopscience.iop.org/article/10.1088/2634-4505/ac84a4}

\bibitem{matzner2019astronomy}
C.~{Matzner}, N.~B. {Cowan}, R.~{Doyon}, V.~{H{\'e}nault-Brunet},
  D.~{Lafreni{\`e}re}, M.~{Lokken}, P.~G. {Martin}, S.~{Morsink},
  M.~{Nomandeau}, N.~{Ouellette}, M.~{Rahman}, J.~{Roediger}, J.~{Taylor},
  R.~{Thacker}, M.~{van Kerkwijk}, {Astronomy in a Low-Carbon Future}, in:
  Canadian Long Range Plan for Astronomy and Astrophysics White Papers, Vol.
  2020, 2019, p.~22.
\newblock \href {http://arxiv.org/abs/1910.01272} {\path{arXiv:1910.01272}},
  \href {http://dx.doi.org/10.5281/zenodo.3758549}
  {\path{doi:10.5281/zenodo.3758549}}.

\bibitem{stevens2019imperative}
A.~R.~H. {Stevens}, S.~{Bellstedt}, P.~J. {Elahi}, M.~T. {Murphy}, {The
  imperative to reduce carbon emissions in astronomy}, Nature Astronomy 4
  (2020) 843--851.
\newblock \href {http://arxiv.org/abs/1912.05834} {\path{arXiv:1912.05834}},
  \href {http://dx.doi.org/10.1038/s41550-020-1169-1}
  {\path{doi:10.1038/s41550-020-1169-1}}.

\bibitem{barret2020estimating}
D.~{Barret}, {Estimating, monitoring and minimizing the travel footprint
  associated with the development of the Athena X-ray Integral Field Unit},
  Experimental Astronomy 49~(3) (2020) 183--216.
\newblock \href {http://arxiv.org/abs/2004.05603} {\path{arXiv:2004.05603}},
  \href {http://dx.doi.org/10.1007/s10686-020-09659-8}
  {\path{doi:10.1007/s10686-020-09659-8}}.

\bibitem{2020NatAs...4..816F}
N.~{Flagey}, K.~{Thronas}, A.~{Petric}, K.~{Withington}, M.~J. {Seidel},
  {Measuring carbon emissions at the Canada-France-Hawaii Telescope}, Nature
  Astronomy 4 (2020) 816--818.
\newblock \href {http://dx.doi.org/10.1038/s41550-020-1190-4}
  {\path{doi:10.1038/s41550-020-1190-4}}.

\bibitem{Jahnke_2020}
K.~Jahnke, C.~Fendt, M.~Fouesneau, I.~Georgiev, T.~Herbst, M.~Kaasinen,
  D.~Kossakowski, J.~Rybizki, M.~Schlecker, G.~Seidel, T.~Henning,
  L.~Kreidberg, H.-W. Rix,
  \href{https://doi.org/10.1038%2Fs41550-020-1202-4}{An astronomical
  institute's perspective on meeting the challenges of the climate crisis},
  Nature Astronomy 4~(9) (2020) 812--815.
\newblock \href {http://dx.doi.org/10.1038/s41550-020-1202-4}
  {\path{doi:10.1038/s41550-020-1202-4}}.
\newline\urlprefix\url{https://doi.org/10.1038%2Fs41550-020-1202-4}

\bibitem{Portegies_Zwart_2020}
S.~P. Zwart, \href{https://doi.org/10.1038%2Fs41550-020-1208-y}{The ecological
  impact of high-performance computing in astrophysics}, Nature Astronomy 4~(9)
  (2020) 819--822.
\newblock \href {http://dx.doi.org/10.1038/s41550-020-1208-y}
  {\path{doi:10.1038/s41550-020-1208-y}}.
\newline\urlprefix\url{https://doi.org/10.1038%2Fs41550-020-1208-y}

\bibitem{Burtscher_2020}
L.~Burtscher, D.~Barret, A.~P. Borkar, V.~Grinberg, K.~Jahnke, S.~Kendrew,
  G.~Maffey, M.~J. McCaughrean,
  \href{https://doi.org/10.1038%2Fs41550-020-1207-z}{The carbon footprint of
  large astronomy meetings}, Nature Astronomy 4~(9) (2020) 823--825.
\newblock \href {http://dx.doi.org/10.1038/s41550-020-1207-z}
  {\path{doi:10.1038/s41550-020-1207-z}}.
\newline\urlprefix\url{https://doi.org/10.1038%2Fs41550-020-1207-z}

\bibitem{Stevens2020}
\href{https://doi.org/10.1038/s41550-020-01216-9}{The climate issue}, Nature
  Astronomy 4~(9) (2020) 811--811.
\newblock \href {http://dx.doi.org/10.1038/s41550-020-01216-9}
  {\path{doi:10.1038/s41550-020-01216-9}}.
\newline\urlprefix\url{https://doi.org/10.1038/s41550-020-01216-9}

\bibitem{2021NatAs...5..857B}
L.~{Burtscher}, H.~{Dalgleish}, D.~{Barret}, T.~{Beuchert}, A.~{Borkar},
  F.~{Cantalloube}, A.~{Frost}, V.~{Grinberg}, N.~{Hurley-Walker},
  V.~{Impellizzeri}, M.~{Isidro}, K.~{Jahnke}, M.~{Willebrands}, {Forging a
  sustainable future for astronomy}, Nature Astronomy 5 (2021) 857--860.
\newblock \href {http://dx.doi.org/10.1038/s41550-021-01486-x}
  {\path{doi:10.1038/s41550-021-01486-x}}.

\bibitem{2021NatAs...5..861A}
A.~{Anderson}, G.~{Maffey}, {Five steps for astronomers to communicate climate
  change effectively}, Nature Astronomy 5 (2021) 861--863.
\newblock \href {http://dx.doi.org/10.1038/s41550-021-01481-2}
  {\path{doi:10.1038/s41550-021-01481-2}}.

\bibitem{Aujoux2021}
C.~Aujoux, K.~Kotera, O.~Blanchard,
  \href{https://linkinghub.elsevier.com/retrieve/pii/S0927650521000311}{Estimating
  the carbon footprint of the {GRAND} project, a multi-decade astrophysics
  experiment}, Astroparticle Physics 131 (2021) 102587.
\newblock \href {http://dx.doi.org/10.1016/j.astropartphys.2021.102587}
  {\path{doi:10.1016/j.astropartphys.2021.102587}}.
\newline\urlprefix\url{https://linkinghub.elsevier.com/retrieve/pii/S0927650521000311}

\bibitem{2021NatRP...3..386A}
C.~{Aujoux}, O.~{Blanchard}, K.~{Kotera}, {How to assess the carbon footprint
  of a large-scale physics project}, Nature Reviews Physics 3~(6) (2021)
  386--387.
\newblock \href {http://arxiv.org/abs/2105.04610} {\path{arXiv:2105.04610}},
  \href {http://dx.doi.org/10.1038/s42254-021-00325-2}
  {\path{doi:10.1038/s42254-021-00325-2}}.

\bibitem{vandertak2021}
F.~van~der Tak, L.~Burtscher, S.~P. Zwart, B.~Tabone, G.~Nelemans, S.~Bloemen,
  A.~Young, R.~Wijnands, A.~Janssen, A.~Schoenmakers,
  \href{https://doi.org/10.1038/s41550-021-01552-4}{{The carbon footprint of
  astronomy research in the Netherlands}}, Nature Astronomy 5~(12) (2021)
  1195--1198.
\newblock \href {http://dx.doi.org/10.1038/s41550-021-01552-4}
  {\path{doi:10.1038/s41550-021-01552-4}}.
\newline\urlprefix\url{https://doi.org/10.1038/s41550-021-01552-4}

\bibitem{2021JATIS...7a7001F}
N.~{Flagey}, K.~{Thronas}, A.~O. {Petric}, K.~{Withington}, M.~J. {Seidel},
  {Estimating carbon emissions at CFHT: a first step toward a more sustainable
  observatory}, Journal of Astronomical Telescopes, Instruments, and Systems 7
  (2021) 017001.
\newblock \href {http://dx.doi.org/10.1117/1.JATIS.7.1.017001}
  {\path{doi:10.1117/1.JATIS.7.1.017001}}.

\bibitem{2022NatAs...6..503K}
J.~{Kn{\"o}dlseder}, S.~{Brau-Nogu{\'e}}, M.~{Coriat}, P.~{Garnier},
  A.~{Hughes}, P.~{Martin}, L.~{Tibaldo}, {Estimate of the carbon footprint of
  astronomical research infrastructures}, Nature Astronomy 6 (2022) 503--513.
\newblock \href {http://arxiv.org/abs/2201.08748} {\path{arXiv:2201.08748}},
  \href {http://dx.doi.org/10.1038/s41550-022-01612-3}
  {\path{doi:10.1038/s41550-022-01612-3}}.

\bibitem{Tsoy2020}
N.~Tsoy, B.~Steubing, C.~van~der Giesen, J.~Guinée,
  \href{https://link.springer.com/10.1007/s11367-020-01796-8}{Upscaling methods
  used in ex ante life cycle assessment of emerging technologies: a review},
  The International Journal of Life Cycle Assessment 25 (2020) 1680--1692.
\newblock \href {http://dx.doi.org/10.1007/s11367-020-01796-8}
  {\path{doi:10.1007/s11367-020-01796-8}}.
\newline\urlprefix\url{https://link.springer.com/10.1007/s11367-020-01796-8}

\bibitem{GRAND20}
{GRAND Collaboration}, J.~{{\'A}lvarez-Mu{\~n}iz}, R.~{Alves Batista},
  A.~{Balagopal V.}, et~al., {The Giant Radio Array for Neutrino Detection
  (GRAND): Science and design}, Science China Physics, Mechanics, and Astronomy
  63~(1) (2020) 219501.
\newblock \href {http://arxiv.org/abs/1810.09994} {\path{arXiv:1810.09994}},
  \href {http://dx.doi.org/10.1007/s11433-018-9385-7}
  {\path{doi:10.1007/s11433-018-9385-7}}.

\bibitem{DONG2021}
L.~Dong, G.~Miao, W.~Wen, China’s carbon neutrality policy: Objectives,
  impacts and paths, East Asian Policy 13 (2021) 5--18.
\newblock \href {http://dx.doi.org/10.1142/S1793930521000015}
  {\path{doi:10.1142/S1793930521000015}}.

\bibitem{COP26}
{United Nations Climate Change Conference UK 2021 in partnership with Italy. },
  Cop 26 the glasgow climate pact (2021). (2021).

\bibitem{DelfUniversityofTechnology}
{Delft University of Technology},
  \href{https://www.ecocostsvalue.com/data/}{{Excel files: Idemat and Ecoinvent
  and ecocosts midpoint tables}} (2021).
\newline\urlprefix\url{https://www.ecocostsvalue.com/data/}

\bibitem{Morgan2012}
R.~K. Morgan,
  \href{http://www.tandfonline.com/doi/abs/10.1080/14615517.2012.661557}{Environmental
  impact assessment: the state of the art}, Impact Assessment and Project
  Appraisal 30 (2012) 5--14.
\newblock \href {http://dx.doi.org/10.1080/14615517.2012.661557}
  {\path{doi:10.1080/14615517.2012.661557}}.
\newline\urlprefix\url{http://www.tandfonline.com/doi/abs/10.1080/14615517.2012.661557}

\bibitem{Payraudeau2005}
S.~Payraudeau, H.~M. van~der Werf,
  \href{https://linkinghub.elsevier.com/retrieve/pii/S0167880905000149}{Environmental
  impact assessment for a farming region: a review of methods}, Agriculture,
  Ecosystems \& Environment 107 (2005) 1--19.
\newblock \href {http://dx.doi.org/10.1016/j.agee.2004.12.012}
  {\path{doi:10.1016/j.agee.2004.12.012}}.
\newline\urlprefix\url{https://linkinghub.elsevier.com/retrieve/pii/S0167880905000149}

\bibitem{Kim2019}
B.~Kim, C.~Azzaro-Pantel, M.~Pietrzak-David, P.~Maussion,
  \href{https://linkinghub.elsevier.com/retrieve/pii/S095965261833186X}{Life
  cycle assessment for a solar energy system based on reuse components for
  developing countries}, Journal of Cleaner Production 208 (2019) 1459--1468.
\newblock \href {http://dx.doi.org/10.1016/j.jclepro.2018.10.169}
  {\path{doi:10.1016/j.jclepro.2018.10.169}}.
\newline\urlprefix\url{https://linkinghub.elsevier.com/retrieve/pii/S095965261833186X}

\bibitem{ISO}
ISO, Environmental management — life cycle assessment — principles and
  framework iso14040:2006 (2006).

\bibitem{ISO2}
ISO, Environmental management — life cycle assessment — requirements and
  guidelines iso14040 (2006).

\bibitem{Chen2021}
X.~Chen, H.~S. Matthews, W.~M. Griffin,
  \href{https://linkinghub.elsevier.com/retrieve/pii/S0921344921002871}{{Uncertainty
  caused by life cycle impact assessment methods: Case studies in process-based
  LCI databases}}, Resources, Conservation and Recycling 172 (2021) 105678.
\newblock \href {http://dx.doi.org/10.1016/j.resconrec.2021.105678}
  {\path{doi:10.1016/j.resconrec.2021.105678}}.
\newline\urlprefix\url{https://linkinghub.elsevier.com/retrieve/pii/S0921344921002871}

\bibitem{Althaus2004}
H.-J. Althaus, G.~Doka, R.~Dones, R.~Hischier, S.~Hellweg, S.~Humbert,
  M.~Margni, T.~Nemecek, M.~Spielmann, \href{www.ecoinvent.ch}{Implementation
  of life cycle impact assessment methods} (5 2004).
\newline\urlprefix\url{www.ecoinvent.ch}

\bibitem{report}
H.~Michael, G.~Jerome, H.~Mark, M.~Manuele, G.~Mark, H.~Reinout, J.~Olivier,
  D.~S. Ab, {Recommendations for Life Cycle Impact Assessment in the European
  context - based on existing environmental impact assessment models and
  factors (International Reference Life Cycle Data System - ILCD handbook)}
  (2011).

\bibitem{Dong2014}
Y.~H. Dong, S.~T. Ng,
  \href{http://link.springer.com/10.1007/s11367-014-0743-0}{{Comparing the
  midpoint and endpoint approaches based on ReCiPe—a study of commercial
  buildings in Hong Kong}}, The International Journal of Life Cycle Assessment
  19 (2014) 1409--1423.
\newblock \href {http://dx.doi.org/10.1007/s11367-014-0743-0}
  {\path{doi:10.1007/s11367-014-0743-0}}.
\newline\urlprefix\url{http://link.springer.com/10.1007/s11367-014-0743-0}

\bibitem{Bare2000}
J.~C. Bare, P.~Hofstetter, D.~W. Pennington, H.~A. {Udo de Haes}, {Life cycle
  impact assessment workshop summary. Midpoints versus endpoints: The
  sacrifices and benefits}, International Journal of Life Cycle Assessment
  5~(6) (2000) 319--326.

\bibitem{Sala2018}
S.~Sala, A.~K. Cerutti, R.~Pant, \href{https://ec.europa.eu/jrc}{{Development
  of a weighting approach for the Environmental Footprint}}, Tech. rep. (2018).
\newblock \href {http://dx.doi.org/10.2760/446145} {\path{doi:10.2760/446145}}.
\newline\urlprefix\url{https://ec.europa.eu/jrc}

\bibitem{Ecoinvent2021}
Ecoinvent, {ecoinvent, Allocation, cut-off by classification, ecoinvent
  database version 3.8} (2021).

\bibitem{Latunussa2016}
C.~E.~L. Latunussa, L.~Mancini, G.~A. Blengini, F.~Ardente, D.~Pennington,
  {Analysis of Material Recovery from Silicon Photovoltaic Panels. Life Cycle
  Assessment and Implications for Critical Raw Materials and Ecodesign}, Tech.
  Rep. March (2016).
\newblock \href {http://dx.doi.org/10.2788/786252} {\path{doi:10.2788/786252}}.

\bibitem{Nassajfar2021}
M.~N. Nassajfar, I.~Deviatkin, V.~Leminen, M.~Horttanainen,
  \href{https://www.mdpi.com/2071-1050/13/21/12126}{{Alternative Materials for
  Printed Circuit Board Production: An Environmental Perspective}},
  Sustainability 13~(21) (2021) 12126.
\newblock \href {http://dx.doi.org/10.3390/su132112126}
  {\path{doi:10.3390/su132112126}}.
\newline\urlprefix\url{https://www.mdpi.com/2071-1050/13/21/12126}

\end{thebibliography}

\end{document}